\pdfoutput=1

\documentclass[prl,twocolumn,showpacs,amsmath,amssymb,floatfix,superscriptaddress]{revtex4-2}
\usepackage{graphicx}
\usepackage{dcolumn}
\usepackage{bm}
\usepackage{amsmath}
\usepackage{braket}
\usepackage{amssymb,xcolor,soul}
\usepackage{pict2e}

\usepackage[normalem]{ulem}

\begin{document}
	
	\title{Elasto-Dynamical Induced Spin and Charge Pumping in Bulk Heavy Metals}
	
	\author{Farzad Mahfouzi}
	\email{Farzad.Mahfouzi@gmail.com}
	\affiliation{Department of Physics and Astronomy, California State University, Northridge, CA, USA}
	\author{Nicholas Kioussis}
	\email{Nick.Kioussis@csun.edu }
	\affiliation{Department of Physics and Astronomy, California State University, Northridge, CA, USA}
	
	\begin{abstract}
	 Analogous to the Spin-Hall Effect (SHE), {\it ab initio} electronic structure calculations reveal that acoustic phonons can induce charge (spin) current flowing along (normal to) its propagation direction. Using Floquet approach we have calculated the elastodynamical-induced charge and spin pumping in bulk Pt and demonstrate that: (i) While the longitudinal charge pumping is an intrinsic observable, the transverse pumped spin-current has an extrinsic origin that depends strongly on the electronic relaxation time; (ii) The longitudinal charge current 
	 is of nonrelativstic origin, while the transverse spin current is a relativistic effect that to lowest order scales linearly with the spin-orbit coupling strength; (iii) both charge and spin pumped currents have parabolic dependence on the amplitude of the elastic wave.
\end{abstract}

	\maketitle
{\it Introduction:}	
One of the primary objectives in the field of spintronics  is the development of efficient means to generate pure  spin current, which can 
be in turn used to manipulate the magnetization configuration and damping rate in magnetic based memory bits\cite{Bhatti2017}, spin transistors\cite{Datta1999,Chuang2015}, antennas\cite{Nan2017}, and sensors\cite{Lenz2006}.
Generating spin current in materials with strong spin-orbit coupling (SOC)
through the spin Hall effect\cite{Dyakonov1971} (SHE), i.e., the transverse spin current generation by electric field,  
has recently been the focus of intensive research both theoretically and experimentally\cite{Manchon2008,Miron2011}. 
The SHE is often parameterized by the charge-to-spin current conversion efficiency, referred to as the spin-Hall angle, $\Theta_{SH}$, 
whereby the spin current is given by $\vec{I}_{\vec{S}}=\frac{\hbar}{2e}\Theta_{SH}\vec{I}_C\times\vec{e}_{\vec{S}}$. 
Here, $\vec{e}_{\vec{S}}$ is the spin polarization unit vector and $\vec{I}_C$ is the charge current. 
Experimentally, the conventional approach to quantify $\Theta_{SH}$  is through spin-orbit torque measurements in ferromagnetic/normal metal bilayer devices.\cite{Hayashi2014,Ghosh2017,Liu2011,Goncalves2013,Kim2013,Mishra2019,Huang2021}

Another commonly used mechanism to generate dc spin current is the 
spin pumping from a precessing ferromagnet (FM) into an adjacent normal metal (NM) \cite{Mahfouzi2012,Suresh2020,Dolui2020,Heinrich2011,Czeschka2011}. 
The generated spin current flowing through the FM/NM interface is ${I}_{\vec{S}}=\frac{\hbar}{4\pi} g_{\uparrow\downarrow}\vec{e}_{\vec{S}}\cdot\vec{m}\times\partial\vec{m}/\partial t$, where $\vec{m}$ and $g_{\uparrow\downarrow}$ denote the direction of the magnetization and the spin-mixing conductance, respectively. $g_{\uparrow\downarrow}$ is often determined experimentally through 
the change of Gilbert damping in the presence and absence of the normal metal adjacent to the FM.\cite{Zhu2019}
In addition, other approaches, such as, the spin polarization effect in magnetic tunnel junctions,\cite{Miyazaki95,Moodera95} the spin Seebeck effect\cite{Uchida2008,Meier2015,Lin2016,Chang2014}, 
spin pumping from magnons excited in response to elastic waves,\cite{Weiler2012,Hayashi2018}
 and spin current pulses produced by ultrafast laser induced demagnetization process  \cite{Hurst2018,Choi2014}
have also been proven to be promising for the generation of spin currents in spintronic devices.

Recently, a different approach to generate spin current was demonstrated experimentally in X/CoFeB/MgO (X=W,Pt,W) heterostructures 
where the spin current emerges from the lattice dynamics  in strong spin-orbit nonmagnetic metals (Pt, W), and flows transverse to the propagation direction of the surface acoustic wave\cite{kawada2021}.  This is similar to the conventional SHE where the spin current propagates orthogonal to the electrical current. To account for the experimental results  Kawada {\it et al.} suggested \cite{kawada2021} that the spin current must scale with the SOC and the time derivative of the lattice displacement along the wave propagation direction.
The suggested plausible mechanisms of the generation of acoustic spin Hall effect include a dynamic change in the Berry curvature of electron wavefunction induced by 
the time-dependent lattice displacement giving rise to a Berry electric field,\cite{niu1999,chaudhary2018} and/or SOC-mediated spin-lattice coupling resembling the Rashba Hamiltonian\cite{romano2008}. Nevertheless, the underlying atomistic mechanism remains unresolved.

\begin{figure}
	\centering
	{\includegraphics[angle=0,trim={0.0cm 6.0cm 2.0cm 5.0cm},clip,width=0.5\textwidth]{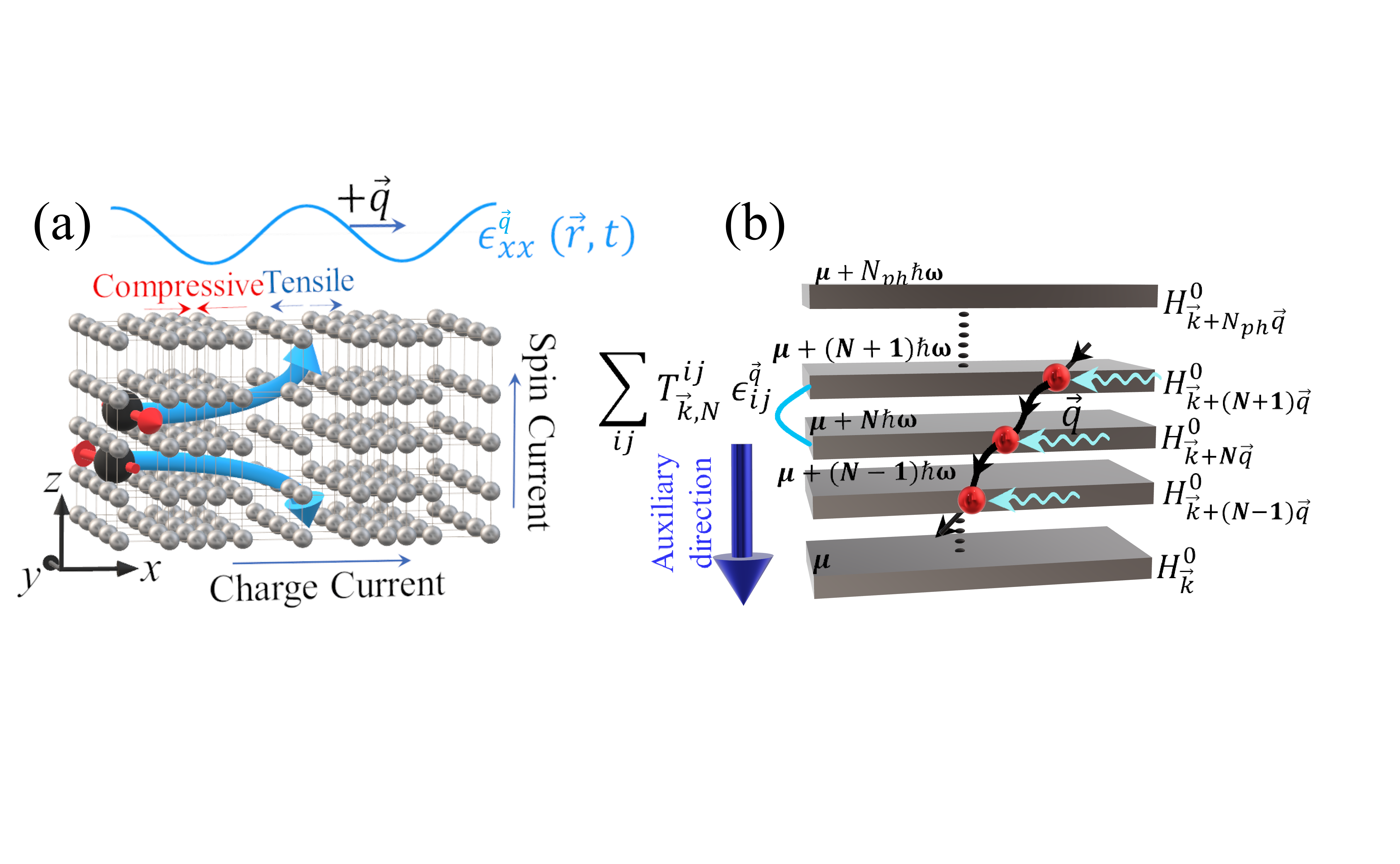}}%
	\caption{ (a) Schematic of the elastodynamical-induced charge and spin current in bulk heavy metal under a time- and position-dependent 
		strain \mbox{$\epsilon_{xx}(\vec{R},t)=Re(\epsilon_{xx}^{\vec{q}}e^{i\omega t+i\vec{q}\cdot\vec{R}})$}, of wavevector $\vec{q}$ along the x direction.  	
The charge current propagates along $\vec{q}$, while the spin current flows (along $z$) orthogonal to the phonon and spin-polarization  (y) directions. 	
(b) Floquet space representation of the combined electron-phonon system. The phononic part consists of a single mode with wave vector $\vec{q}$ and a cutoff number, $N_{ph}$, of phonons. The $N$th layer represents the electronic system with Hamiltonian $H^0_{\vec{k}+N\vec{q}}$ and chemical potential $\mu+N\hbar\omega$ that is coupled to the other electronic systems with different number of phonons though the electron-phonon coupling matrix, $T^{ij}_{\vec{k},N}$, $i,j=x,y,z$.}
	\label{fig:fig1}
\end{figure}

In this Letter, using {\it ab initio} based electronic structure calculations we reveal the emergence of a dc charge (spin) current in response to acoustic phonons in heavy metals (Pt), shown schematically in Fig. \ref{fig:fig1}(a),
where the spin current flows transverse to the phonon propagation direction.  
We demonstrate that the phonon-induced spin current is a relativistic effect arising from the spin-orbit coupling (SOC) where, to the lowest order depends linearly on the SOC strength. Analogous to the SHE, the spin polarization orientation is orthogonal to both the spin current and phonon propagation directions. 
We show that the phonon-induced charge (spin) current  saturates (diverges) in the limit of ballistic transport regime, signifying its intrinsic (extrinsic) origin. This is in sharp contrast to the electric field-induced charge (spin) current where the corresponding longitudinal (transverse) conductivity is extrinsic (intrinsic).

{\it Theoretical Formalism:} In the linear response regime the electronic Hamiltonian of a bulk system under a
time- and position-dependent strain, $\epsilon_{ij}(\vec{R},t)$ ($i,j = x, y, z$), is given by\cite{Klimeck2014,Boykin2002} 
\begin{align}\label{eq:Ham1}
	&\hat{\bf H}=\hat{\bf H}^0+\frac{1}{2}\sum_{ij}(\hat{\bf H}^{ij}\hat{\boldsymbol{\epsilon}}_{ij}(t)+\hat{\boldsymbol{\epsilon}}_{ij}(t)\hat{\bf H}^{ij}).
\end{align}
Here, bold symbols denote matrices in real space, hat symbols denote matrices in atomic orbital Hilbert space, $\hat{\bf H}^0$ is the Hamiltonian in the absence of strain, ${\hat{\bf H}^{ij}} = \partial\hat{\bf H}/\partial\epsilon_{ij}|_{\epsilon_{ij}\rightarrow0}$ 
is the deformation Hamiltonian term associated with the coupling between electrons and local strain,
$[\hat{\bf H}]_{\vec{R},\vec{R}'}=\hat{H}_{\vec{R}-\vec{R}'}$, 
$\vec{R},\vec{R}'$ are the positions of the unit cells, and the time- and position-dependent local strain is a diagonal matrix with elements given by, $[{\hat{\boldsymbol{\epsilon}}}_{ij}]_{\vec{R},\vec{R}'}=\hat{1}{\epsilon}_{ij}(\vec{R},t)\delta_{\vec{R},\vec{R}'}$.  Eq.~\eqref{eq:Ham1} assumes that spatial variation of strain is adiabatic and hence ignores the dependence of the Hamiltonian on the gradience of strain.
For a single phonon mode of wave vector, $\vec{q}$, and frequency, $\omega$, we have,
 \mbox{$\epsilon_{ij}(\vec{R},t)=Re(\epsilon^{\vec{q}}_{ij}e^{i\omega t+i\vec{q}\cdot\vec{R}})$}. In this case,  
 the time and position dependence of the Hamiltonian can be removed by applying a gauge transformation, $[\hat{\bf U}]_{NM}=\delta_{NM}\hat{1}e^{iN\omega t+iN\vec{q}\cdot\vec{R}}$,  where the capital letters, N and M refer to phonon states.
 The resulting time-independent system is referred to as the Floquet space, where the corresponding Hilbert space is extended to include 
 the phononic degrees of freedom, with the total Hamiltonian given by 
\begin{align}\label{eq:Ham11}
	[\hat{\mathcal H}_{\vec{k}}]_{NM}&=\hat{H}^0_{\vec{k}+N\vec{q}}\delta_{NM} \nonumber\\
	&+\frac{1}{2}\sum_{ij}\hat{T}^{ij}_{\vec{k},N}\epsilon^{\vec{q}}_{ij}\Big(\delta_{N,M+1}+\delta_{N,M-1}\Big),
\end{align}
where $\hat{T}^{ij}_{\vec{k},N}= \frac{1}{2} 
\big(\hat{H}^{ij}_{\vec{k}+N\vec{q}}+\hat{H}^{ij}_{\vec{k}+(N+1)\vec{q}}\big)$, $N, M = 0, \ldots, N_{ph}$, and $N_{ph}$ is the cutoff for the number of phonons. 
The hybrid electron-phonon Hamiltonian is shown schematically in Fig.~\ref{fig:fig1}(b), where the additional dimension is introduced to account for the single mode phononic degree of freedom. The dependence of the quasi-particle chemical potential on the number of phonons results in a quasiparticle transport along the auxiliary direction which describes the attenuation of the single mode elastic wave through electron-phonon scattering.\cite{Mahfouzi2017} In this case, due to the conservation of momentum, both electrons and holes are dragged along the phonon propagation direction, whereby systems with electron-hole asymmetry experience a net pumped charge current. This mechanism of pumping is, in essence, similar to the phonon drag effect that gives rise to an enhancement of Seebeck coefficient which is often treated  theoretically using coupled electron-phonon Boltzmann transport equations.\cite{Zhou2015}  

For an isolated  system, the wave function of the coupled electron-phonon system is of the form, $|N\alpha\vec{k}\rangle=|N\rangle\otimes |\alpha\vec{k}\rangle$, where, $\otimes$ refers to the Kronecker product, and $\alpha$ denotes the atomic orbitals and spin of the electron Bloch states. The single-quasi-particle retarded Green function and the corresponding density matrix can be obtained from,\cite{Mahfouzi2017,Mahfouzi2017b}
\begin{align}\label{eq:GFeq2}
	&\hat{\boldsymbol{\rho}}_{\vec{k}}=\frac{\eta}{\pi} \int dE\hat{\boldsymbol{G}}_{\vec{k}}(E){f}(E\hat{\bf 1}-\hat{\boldsymbol{\Omega}})\hat{\boldsymbol{G}}_{\vec{k}}^{\dagger}(E),
\end{align}
where the Green's function is calculated from,
\begin{align}\label{eq:GFeq1}
	&\left(E-i\eta-\hat{\bf \mathcal{H}}_{\vec{k}}-\hat{\boldsymbol{\Omega}}\right)\hat{\boldsymbol{G}}_{\vec{k}}(E)=\hat{\boldsymbol{1}}.
\end{align}
Here, $[\hat{\boldsymbol{\Omega}}]_{NM}=N\hbar\omega\delta_{NM}\hat{1}$, is the single mode phononic Hamiltonian, $f(E)$ is the Fermi-Dirac distribution function and $\eta=\hbar/2\tau$ is the energy broadening parameter which is inversely proportional to the electronic relaxation time, $\tau$. Using density matrix  given by Eq.~\eqref{eq:GFeq1}, the charge and spin currents are determined from, 
\begin{subequations}\label{eq:Ieq}
	\begin{align}
		\vec{I}&=e\langle \vec{\hat{\boldsymbol{v}}}_{\vec{k}}\rangle,\label{eq:Ieqa}\\
		\vec{I}^{S_i}&= \frac{\hbar}{2}Re\langle\hat{\sigma}_i\vec{\hat{\boldsymbol{v}}}_{\vec{k}}\rangle,\label{eq:Ieqb}	
	\end{align}
\end{subequations}
where, $\langle...\rangle=\sum_{\vec{k}}Tr[...\hat{\boldsymbol{\rho}}_{\vec{k}}]/(VN_kN_{ph})$ is the expectation value, $V$ is the unit of the unit cell, $N_k$ is the number of k-points in the summation and $\vec{\hat{\boldsymbol{v}}}_{\vec{k}}=\partial\hat{\mathcal H}_{\vec{k}}/\partial \vec{k}$ is the electronic group velocity operator.  In the ballistic regime and in linear response to the phonon frequency, $\omega$, the density matrix is given by,
\begin{align}\label{eq:DensMateq2}
	&\hat{\boldsymbol{\rho}}_{\vec{k}}\approx f(\mathcal{\hat{H}}_{\vec{k}}+\hat{\boldsymbol{\Omega}})+\frac{\eta }{\pi}\hat{\boldsymbol{G}}_{\vec{k}}(E_F)\hat{\boldsymbol{\Omega}}\hat{\boldsymbol{G}}_{\vec{k}}^{\dagger}(E_F) .
\end{align}
The first and second terms in Eq.~\eqref{eq:DensMateq2} are referred to as the Fermi sea and Fermi surface contributions, respectively.
The Fermi surface contribution to the density matrix  and the resulting pumped charge and spin currents can be separated into even and odd components with respect to $\eta$ (or relaxation time), $\hat{\boldsymbol{\rho}}^{in/ex}_{\vec{k}}=(\hat{\boldsymbol{\rho}}_{\vec{k}}(\eta)\pm\hat{\boldsymbol{\rho}}_{\vec{k}}(-\eta))/2$. To lowest order, the even (odd) component is expected to be independent of (inversely proportional to) $\eta$, demonstrating its intrinsic (extrinsic) character. The term ``extrinsic" is defined as the odd component of the density matrix with respect to the broadening parameter $\eta$, which, to lowest order, is inversely proportional to $\eta$ and originates from the intraband contributions (See Supplemental Material \cite{SM}). 

In the ballistic regime, $\eta\rightarrow 0$, the intrinsic (even) component of the charge current can be written in terms of the Berry curvature, 

\begin{align}
	\vec{I}^{in}&=\frac{2e}{VN_kN_{ph}}\sum_{n\vec{k}} f(\varepsilon_{n\vec{k}})
	Im\bra{\frac{\partial\psi_{n\vec{k}}}{\partial\vec{k}}}\hat{\boldsymbol{U}}^{\dagger}\frac{\partial}{\partial t}\boldsymbol{\hat{U}}\ket{\psi_{n\vec{k}}}\nonumber\\
	&=\frac{2e}{VN_kN_{ph}}\sum_{n\vec{k}} f(\varepsilon_{n\vec{k}}) Re\bra{\frac{\partial\psi_{n\vec{k}}}{\partial\vec{k}}}\boldsymbol{\hat{\Omega}}\ket{\psi_{n\vec{k}}},\label{eq:Berryeq2}
\end{align}
where, we use perturbation theory to express, \mbox{$\ket{\frac{\partial\psi_{n\vec{k}}}{\partial\vec{k}}}=\sum_{m}\frac{(\vec{\boldsymbol{\hat{v}}})_{nm}\ket{\psi_{m\vec{k}}}}{\varepsilon_{n\vec{k}}-\varepsilon_{m\vec{k}}-i\eta}$}, with $\vec{\boldsymbol{\hat{v}}}_{nm,\vec{k}}=\bra{\psi_{n\vec{k}}}\vec{\boldsymbol{\hat{v}}}\ket{\psi_{m\vec{k}}}$ being
the group velocity matrix elements,  and the energy broadening parameter, $\eta$, is introduced to avoid divergences at the degenerate points. The wavefunction, $\ket{\psi_{n\vec{k}}}$ and energy dispersion, $\varepsilon_{n\vec{k}}$ are the eigenvectors and eigenvalues of the Floquet Hamiltonian, $\hat{\bf \mathcal{H}}_{\vec{k}}$, respectively. 
The intrinsic component of the pumped spin current can be calculated from an expression similar to 
Eq.~\eqref{eq:Berryeq2}, by replacing the charge current operator, $e\vec{\boldsymbol{\hat{v}}}$, with the spin current operator  given by the Hermitian part of $\hbar\hat{\sigma}_i\vec{\boldsymbol{\hat{v}}}/2$.  Furthermore, it should be noted that in systems with small bandgap at the Fermi surface, topological nature of Eq.~\eqref{eq:Berryeq2} can yield quantized values (i.e. independent of elastic wave amplitude) where  electrons and holes are trapped and dragged by the elastic wave,  resulting in the so-called Thouless pumping.\cite{Thouless1983,Leek2005}

\begin{table}[t]
	\centering
	\caption{Values of the intrinsic and extrinsic contributions to the
	elastodynamical-induced components	
	of the	pumped charge, $\vec{I}$, and spin current,   $\vec{I}^{S_{i}}$, divided by the acoustic phonon frequency, $\omega$,
	(in units of $10^{10}/e\Omega m^2$), for phonon wavevector,  
		$\vec{q}= (\pm 0.01,0,0)$(\AA$^{-1}$) and strain amplitude $\epsilon^{\vec{q}}_{xx}=10^{-3}$.}
	\begin{ruledtabular}
		\begin{tabular}{ c || c c c || c c c }
			&    &  Intrinsic (Even)  &       &       & Extrinsic (Odd)  &     \\
			\hline
			$i =$              &$x$        &   $y$   &  $z$   & x  & y  & z \\
			\hline
			$I_{i}(\vec{q})$/$\omega$  & $\pm$80   &0       &0      &0   & 0  & 0 \\
			$I^{S_x}_{i}(\vec{q})$/$\omega$                          &0         &0       &0      &0   &0  &0       \\ 
			$I^{S_y}_{i}(\vec{q})$/$\omega$                          &0         &0       &0      &0   &0  & $\mp$ 0.7       \\ 
			$I^{S_z}_{i}(\vec{q})$/$\omega$                          &0         &0       &0      &0   &$\pm$0.7  & 0       \\ 	
		\end{tabular}
	\end{ruledtabular}
	\label{tab:my_label}
\end{table}
{\it Computational Approach:}
The tight-binding Hamiltonian, $\hat{H}_{\vec{k}}$ matrices for bulk Pt under different values of strain, $\epsilon_{ij}$, 
are calculated using the Linear Combination of Atomic Orbitals OpenMX package\cite{OzakiPRB2003,OzakiPRB2004,OzakiPRB2005}.
For a nonmagnetic material, the one-electron Kohn-Sham Hamiltonian can be expressed by\cite{Mahfouzi2017,Mahfouzi2018,Mahfouzi2020}, 
\begin{align}\label{eq:Hamil2}
	\hat{H}_{\vec{k}}(\{\epsilon_{ij}\})&=\hat{H}_{\vec{k}}^K(\{\epsilon_{ij}\})\hat{1}_{2\times 2}+\xi\hat{H}^{soc}_{\vec{k}}(\{\epsilon_{ij}\}),
\end{align}
where, the first term represents the kinetic component of the Hamiltonian, the second term is the SOC contribution, and $\xi$ is the scaling factor. 	
The effect of strain, ${\epsilon_{ij}}$, is to modify the primitive lattice vectors, $\vec{a}'_i$, such that, $(\vec{a}'_i-\vec{a}_i)\cdot\vec{e}_j=\sum_{k}\vec{a}_{i}\cdot\vec{e}_k\epsilon_{kj}$, where the  $\vec{e}_j$'s denote unit vectors in Cartesian coordinates. 
The electron-phonon coupling terms in Eq.~\eqref{eq:Ham1} are calculated by fitting Eq.~\eqref{eq:Hamil2} to a polynomial function, and in turn calculate, $\hat{H}^{ij}_{\vec{k}}=\partial\hat{H}_{\vec{k}}/\partial\epsilon_{ij}|_{\epsilon_{ij}\rightarrow0}$.
The self-consistent (SCF) calculations employ the Troullier-Martins type norm-conserving pseudopotentials\cite{TroullierPRB1991} with partial core correction. 
The equilibrium lattice constant of bulk Pt was set to  $a=3.96$ \AA. 
In the SCF calculations we used a $24^3$ k-point mesh in the first Brillouin zone, and an energy cutoff of 350 Ry for numerical integrations in the real space grid. 
For the exchange correlation functional the LSDA\cite{CeperleyPRL1980} parameterized by Perdew and Wang\cite{PerdewPRB1981} was used.

\begin{figure}
	\includegraphics[scale=0.25,angle=0,trim={0cm 7cm 0.0cm 6.5cm},clip,width=0.5\textwidth]{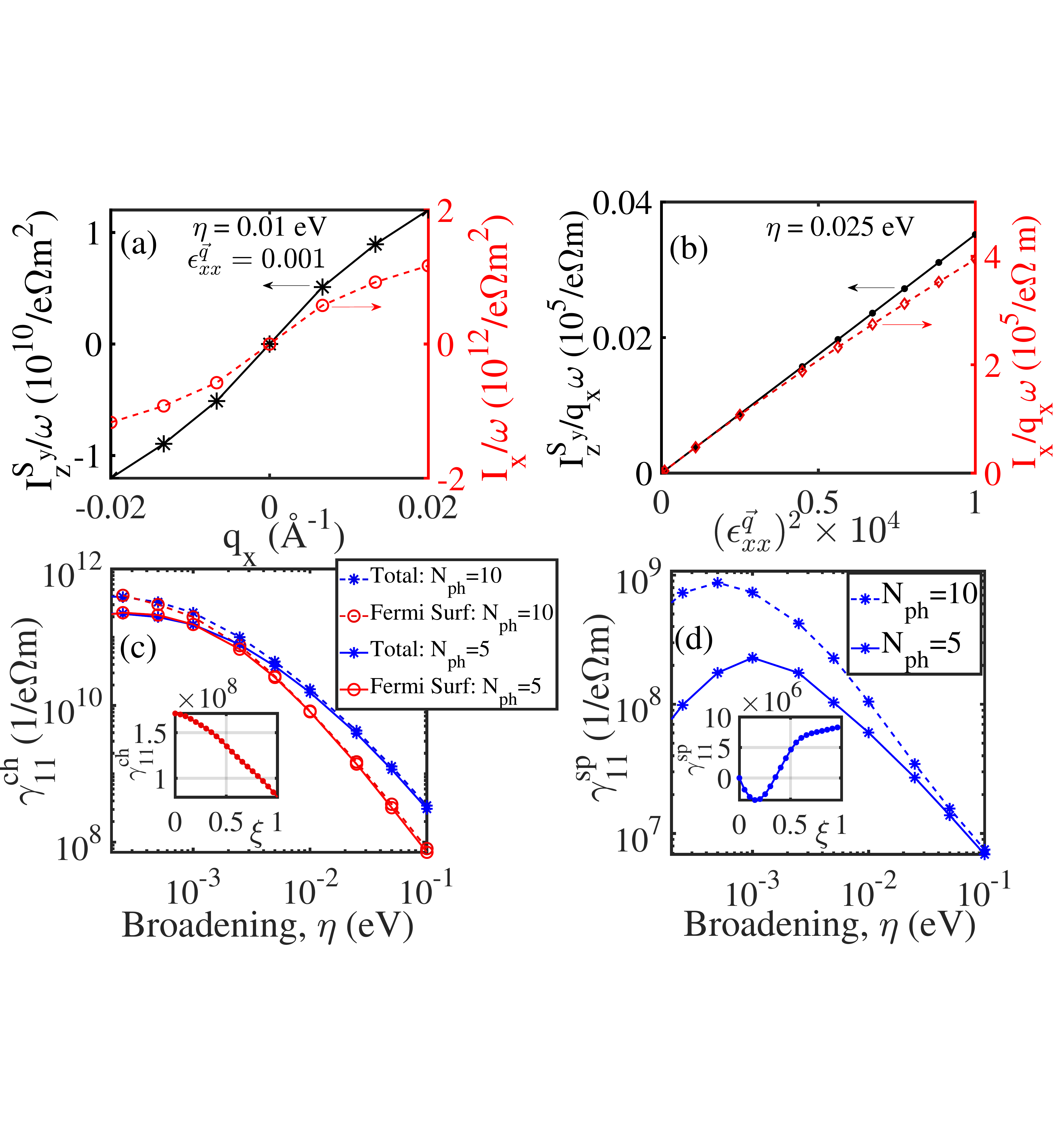}
	\caption{(a) Elastodynamical pumped spin (left-hand ordinate) and charge (right-hand ordinate) current divided by the phonon frequency, $\omega$, versus the phonon wavevector $\vec{q}$ along $x$-axis with strain amplitude $\epsilon^{\vec{q}}_{xx}=10^{-3}$ and energy broadening parameter $\eta$= 10 meV. 
	(b) Efficiency of the extrinsic elastodynamical-induced spin 
    current, $I^{S_y;ex}_z/q_x\omega$ , (left-hand ordinate) and intrinsic charge current, $I^{in}_x/q_x\omega$, (right-hand ordinate) in the limit $q_x \rightarrow$ 0, versus the square of the strain amplitude, $(\epsilon^{\vec{q}})^2_{xx}$, where $\epsilon^{\vec{q}}_{xx}$ ranges between 0 and 1$\%$. 
	(c) Total (stars) and Fermi surface (circles) contributions to the Elastodynamical Longitudinal Charge Conductivity (ELCC) versus $\eta$ 
	for $N_{ph}$= 5 and 10 shown with solid and dashed curves, respectively. 
	(d) Fermi surface contribution to the Elastodynamical Transverse Spin Conductivity (ETSC) versus $\eta$ for different values of $N_{ph}$. 
	Insets in (c) and (d):  ELCC and ETSC versus SOC scaling factor or $\eta=0.1$ eV.}
	\label{fig:fig2}
\end{figure}

{\it Results and Discussion:} Table I lists the values of the intrinsic and extrinsic contributions to the elastodynamical-induced components	
of the	pumped charge, $\vec{I}$, and spin current,  $\vec{I}^{S_{i}}$, divided by the acoustic phonon frequency, $\omega$, (in units of $10^{10}/e\Omega m^2$), for   
$\vec{q}=(\pm 0.01,0,0)$(\AA$^{-1}$), strain amplitude $\epsilon^{\vec{q}}_{xx}=10^{-3}$ and 
$\eta$ = 0.01eV. The charge current is dominated by the intrinsic component (even function of $\eta$) and is parallel to the elastic wave propagation direction, $\vec{I}\parallel\vec{q}$, while the spin current is extrinsic (odd function of $\eta$) and propagates normal to $\vec{q}$ and the spin polarization direction, $\vec{I}^{\vec{S}}\parallel\vec{e}_{\vec{S}}\times\vec{q}$.
Fig.~\ref{fig:fig2}(a) displays the calculated pumped charge (dashed curve) and spin (solid curve) currents versus the wavevector of the elastic wave propagating along $x$, respectively, with $\epsilon^{\vec{q}}_{xx}$ = $10^{-3}$ and $\eta$= 0.01eV.
Both charge and spin currents are odd functions of the wavevector with a linear dependence on $q_x$ close to the $\Gamma$-point.  This is analogous to the electric field-induced charge and spin current, where the external electric field is replaced with $\hbar\omega\vec{q}/e$. The difference, however, is that in contrast to the extrinsic (intrinsic) nature of the Ohmic (spin-Hall) current, the phonon-induced charge (spin) current exhibits intrinsic (extrinsic) character.


The slope of the pumped charge and spin currents with respect to $q_x$ in the limit of $q_x\rightarrow 0$, describes the efficiency of the phonon-induced electronic transport in the material. Hence, in Fig.~\ref{fig:fig2}(b) we show the variation of the elastodynamical-induced even  (intrinsic) component of the 
charge current, $I^{in}_x/q_x\omega$, and odd  (extrinsic) component of the  spin current, $I^{S_y;ex}_z/q_x\omega$ , 
versus $(\epsilon^{\vec{q}}_{xx})^2$. The calculations reveal that both the charge (dashed curve) 
and the spin current (solid curve) vary quadrtiaclly with the amplitude of the elastic wave for a wide range of strain.
These results suggest that the  strain dependence of the elastodynamical-induced pumped charge and spin currents can be written in the general form, 
\begin{subequations}
	\begin{align}
		\vec{I}({\vec{q}})/\omega&=\vec{q}\sum_{i\le j,k\le l}\gamma^{ch}_{ij,kl}\epsilon^{\vec{q}}_{ij}\epsilon^{\vec{q}}_{kl}\label{eq:9a}\\
		\vec{I}^{\vec{S}}({\vec{q}})/\omega&=\frac{\hbar}{2e}\vec{e}_{\vec{S}}\times\vec{q}\sum_{i\le j,k\le l}\gamma^{sp}_{ij,kl}\epsilon^{\vec{q}}_{ij}\epsilon^{\vec{q}}_{kl},\label{eq:9b}
	\end{align}
\end{subequations}
where, $\gamma^{ch}_{ij,kl}$ and $\gamma^{sp}_{ij,kl}$ are the Elastodynamical Longitudinal Charge Conductivity (ELCC) and Elastodynamical Transverse Spin Conductivity (ETSC), respectively. Given,  $\epsilon_{ij}=\epsilon_{ji}$, in order to avoid double counting we consider only strains with, $i\le j$. One can then use the Voigt notation ($[1,2,3,4,5,6]\equiv [xx,yy,zz,yz,xz,xy]$) to represent the $\gamma_{ij,kl}^{ch/sp}$ tensor elements as a 6$\times$6 symmetric  matrix.

The variation of the ELCC versus the broadening energy, $\eta$, is shown in Fig.~\ref{fig:fig2}(c) for phonon cutoff number, $N_{ph}=$ 5 and 10. We display the variation of the total ELCC and its Fermi surface contribution, where the former is determined  from Eq.~\eqref{eq:Berryeq2}, while the latter is calculated   
from Eq.~\eqref{eq:Ieqa} where the density matrix is given only by the second term in Eq.~\eqref{eq:DensMateq2}. 
 We find that in the limit of ballistic regime ($\eta \rightarrow$ 0), the ELCC saturates to a finite value consistent with the expected behavior of an intrinsic observable. On the other hand, in the limit of large $\eta$ (diffusive regime), while the ELCC becomes independent of $N_{ph}$, the deviation between the total and Fermi surface contribution becomes more significant, 
 due to the fact that Eq.~\eqref{eq:Berryeq2} is derived and hence valid in the small $\eta$ limit. The inset in 
Fig.~\ref{fig:fig2}(c) shows the dependence of ELCC on the SOC scaling factor with a finite value in the absence of SOC, demonstrating its non-relativistic nature. 

Fig.~\ref{fig:fig2}(d) shows the variation of the extrinsic component of ETSC (only the Fermi surface contribution) versus $\eta$ for $N_{ph}=$ 5 and 10. In contrast to the expected $1/\eta$ dependence, we find that in the limit of $\eta\rightarrow 0$ the ETSC is proportional to $\eta$ and reaches a peak with increasing  $\eta$. 
The peak ETSC value increases with $N_{ph}$, and the corresponding $\eta_{max}$  decreases with  increasing $N_{ph}$, suggesting that ETSC$\propto\eta/(N^{-1}_{ph}+c\eta^2)$, where $c$ is a constant. This means, a larger phonon cutoff number is required to converge the ETSC value in  the clean limit (small $\eta$ limit). 
The inset in Fig.~\ref{fig:fig2}(d) shows the ETSC versus the SOC scaling factor, $\xi$, demonstrating its relativistic nature, which, analogous to the SHE, to  lowest order in $\xi$, is proportional to the SOC strength. It is worth noting that the non-monotonic dependence of ETSC and its sign reversal with SOC for small $\eta$ suggests that in heavy metals (with large SOC) such as Pt a perturbative treatment of ETSC with respect to SOC may fail. Interestingly, the results of ETSC versus SOC shown in Fig. S1 in the supplemental material\cite{SM} for larger $\eta$ of 0.5 eV exhibits a monotonic behavior. 

Thus far, we have focused only on the elastodynamical-induced pumping in response to longitudinal elastic wave, $\epsilon_{xx}(\vec{R},t)$. 
In the following we  present the results for the non-zero matrix elements of the  6$\times$6 elastodynamical charge/spin conductivity matrix.  
The values of the diagonal matrix elements for the even (intrinsic) component of the ELCC matrix (in units of $10^9/e\Omega m$) for $\vec{q}\parallel x$ and  $\eta=25\ meV$, are  $\gamma^{ch}_{ii}= [1.5, 2.5, 2.5, 0.8, 1,  1]$, ($i$=1, $\ldots$, 6), while those of the off-diagonal matrix elements are $\gamma^{ch}_{12}=\gamma^{ch}_{13}=-0.6$, and  $\gamma^{ch}_{23}=-2.4$, where $\gamma^{ch/sp}_{ij}=\gamma^{ch/sp}_{ji}$.
Note that similar to the elastic stiffness and magneto-elastic tensor elements\cite{Mahfouzi2020b}, the symmetry of the crystal structure for the elastodynamical electronic transport reduces the number of independent tensor elements. However, the difference is that the elastic wave propagation along $x$-axis breaks the cubic symmetry and the ELCC matrix elements resemble those of a tetragonal system instead.

Similar calculations for the extrinsic component of the ETSC (in units of $10^7/e\Omega m$) with spin current along the $z$-axis and spin-polarization along $y$-axis, yields that the nonzero diagonal elements are, $\gamma^{sp}_{ii}=[3.5, 2.6, -0.6, 2, 0.5,-0.4]$ and the off diagonal elements are, $\gamma^{sp}_{12}=-3.3$, $\gamma^{sp}_{13}=-1.6$ and $\gamma^{sp}_{23}=1.6$. Note that all of the nonzero elements of ETSC are independent, similar to the elastic stiffness matrix elements of an orthorhombic crystal structure. This is due to the fact that the choice of the spin current direction along $z$ renders the 
$yz$-plane ({\it i.e.,} normal to $\vec{q}\parallel\vec{e}_x$) anisotropic. 

As demonstrated in the Supplemental Material\cite{SM}, the underlying mechanism of the phonon-induced pumping can be attributed to the electron-hole asymmetry, which
results in a negative (positive) sign for ELCC when the transport is electron-(hole-)like. 
The derivative of density of states (DOS) with respect to the energy  can be employed to qualitatively determine the sign of the phonon-induced charge pumping. 
We find a correlation between ELCC, ETSC and the derivative of DOS as a function of chemical potential shift [see Fig. S2 in \cite{SM}],
 suggesting that the ETSC may be explained by the phonon drag charge current combined with the spin-Hall effect.


An approximate value of the experimental ELCC can be determined from 
$\vec{I} \approx \omega\vec{q}\gamma^{ch}_{eff}\epsilon_{eff}^2$, using the measured values\cite{kawada2021} of pumped charge current, $I=7.5\times10^{3} (A/m^2)$, in response to an elastic wave with frequency $\nu=$193 MHz, wavelength, $\lambda$= 20 $\mu m$, and strain amplitude $\epsilon_{eff}\approx 0.01\%$.\cite{Nakano1997} 
An effective value of ELCC, $\gamma^{ch,exp}_{eff}\approx 10^{12} (1/e\Omega m)$ is then estimated, which is near the upper limit of the theoretical value in Fig.~\ref{fig:fig2}(c), corresponding to the clean limit, $\eta\rightarrow 0$.
A more accurate comparison with experiment requires taking into account all  ELCC tensor elements. 
Moreover, the spin current calculated in this work accounts only for the conventional bulk component and neglects the spin-torque \cite{Shi2006,Zhang2008} and the FM/NM interfacial \cite{Mahfouzi2020} contributions. Therefore, a proper comparison with experiment requires calculations of the elastodynamical spin orbit torque in bilayer systems. It should also be noted that the relaxation time approximation is reliable only in the small $\eta$ limit, since the limit of large $\eta$ violates conservation laws, resulting in a smaller pumped spin-current.\cite{Baym1961,Mahfouzi2017}

In conclusion, we have developed a Floquet-based approach using density functional theory and demonstrated the emergence of charge and spin currents induced by an acoustic phonon in bulk Pt. The calculations unveil the underlying atomistic mechanism of the recently discovered acoustic spin Hall effect in strong spin-orbit metals.\cite{kawada2021}
We find that the pumped charge (spin) current is intrinsic (extrinsic) property,  flows along (normal to) to the phonon wavevector, and is of nonrelativistic (relativistic) origin. It is worth mentioning that the phonon-induced pumped charge and spin current effect presented in this Letter can be generalized to other bosonic excitations (e.g. magnons, photons, etc) and their combinations.  

The authors are grateful for fruitful discussions with Masamitsu Hayashi and Takuya Kawada. The work is supported by NSF ERC-Translational Applications of Nanoscale Multiferroic Systems (TANMS)- Grant No. 1160504, NSF PFI-RP Grant No. 1919109,
and by NSF-Partnership in Research and Education in Materials (PREM) Grant No. 
DMR-1828019.



\end{document}